\documentclass[twocolumn,prl,showpacs,aps]{revtex4-1}

\usepackage[english]{babel}
\usepackage{times}
\usepackage{graphicx}
\usepackage{graphics}
\usepackage{amsmath}
\usepackage{amsfonts}
\usepackage{amssymb}
\usepackage{dsfont}
\usepackage{epstopdf}
\usepackage{makeidx}
\usepackage{subfigure}
\usepackage{color}
\usepackage{pgf}
\usepackage{bm}
\usepackage{tikz} 
\usepackage[normalem]{ulem}

\newcommand{\be}{\begin{equation}}
\newcommand{\ee}{\end{equation}}
\newcommand{\bea}{\begin{eqnarray}}
\newcommand{\eea}{\end{eqnarray}}

\makeindex
\begin{document}
\title{Tuning quantum fluctuations with an external magnetic field: Casimir-Polder interaction between an atom and a graphene sheet}
\author{T. Cysne*}
\affiliation{Instituto de F\'{\i}sica, Universidade Federal do Rio de Janeiro,
Caixa Postal 68528, Rio de Janeiro 21941-972, RJ, Brazil}

\author{W. J. M. Kort-Kamp*}
\affiliation{Instituto de F\'{\i}sica, Universidade Federal do Rio de Janeiro,
Caixa Postal 68528, Rio de Janeiro 21941-972, RJ, Brazil}

\author{D. Oliver}
\affiliation{Instituto de F\'{\i}sica, Universidade Federal do Rio de Janeiro,
Caixa Postal 68528, Rio de Janeiro 21941-972, RJ, Brazil}
\author{F. A. Pinheiro}
\affiliation{Instituto de F\'{\i}sica, Universidade Federal do Rio de Janeiro,
Caixa Postal 68528, Rio de Janeiro 21941-972, RJ, Brazil}
\author{F. S. S. Rosa}
\affiliation{Instituto de F\'{\i}sica, Universidade Federal do Rio de Janeiro,
Caixa Postal 68528, Rio de Janeiro 21941-972, RJ, Brazil}
\author{C. Farina}
\affiliation{Instituto de F\'{\i}sica, Universidade Federal do Rio de Janeiro,
Caixa Postal 68528, Rio de Janeiro 21941-972, RJ, Brazil}
\date{\today}

\begin{abstract}
We investigate the dispersive Casimir-Polder interaction between a Rubidium atom and a suspended graphene sheet subjected to an external magnetic field ${\bf B}$. We demonstrate that this concrete physical system allows for an unprecedented control of dispersive interactions at micro and nanoscales. Indeed, we show that the application of an external magnetic field can induce a $80\%$ reduction of the Casimir-Polder energy relative to its value without the field. We also show that sharp discontinuities emerge in the Casimir-Polder interaction energy for certain values of the applied magnetic field at low temperatures. Moreover, for sufficiently large distances these discontinuities show up as a plateau-like pattern with a quantized Casimir-Polder interaction energy, in a phenomenon that can be explained in terms of the quantum Hall effect. In addition, we point out the importance of thermal effects in the Casimir-Polder interaction, which we show that must be taken into account even for considerably short distances. In this case, the discontinuities in the atom-graphene dispersive interaction do not occur, which by no means prevents the tuning of the interaction in $\sim 50 \%$ by the application of the external magnetic field.
\end{abstract}
\maketitle
{\footnotesize *These authors contributed equally to this work and are joint first authors.}
%

\vspace{0.1in}

It has been known for a long time that quantum fluctuations give rise to
interactions between neutral but polarizable objects (atoms, molecules or even
macroscopic bodies) which do not possess any permanent electric or magnetic
multipoles. These are referred to as dispersive interactions, first explained in
the non-retarded regime by Eisenchitz and London \cite{London-Eisenchitz-30}.
Retardation effects were first reported by Casimir and Polder
\cite{Casimir-Polder-46, Casimir-Polder-48} in works that pioneered in the study
of dispersive interactions between an atom and a perfectly conducting plane for
arbitrary distances, generalizing the Lennard-Jones non-retarded result
\cite{Lennard-Jones}. Since then these interactions, nowadays known as
Casimir-Polder forces, have been diligently investigated both theoretically
\cite{Buhmann08,BuhmannLivro,Milonni,CasimirLivro,MostepanenkoRMP,MostepanenkoLivro, Thiru} and experimentally \cite{Druzhinina,Laliotis,Obrecht,Pasquini,Shimizu,Sukenik,Sukenik,Landragin}. If
instead perfectly conducting plates one considers dispersive media, the
calculations become more involved; this has motivated the development of a
general theory of dispersive forces, including thermal effects
\cite{Lifshitz-56, DLP}. Dispersive forces play an important role
in many different areas of research and applications, ranging from biology and
chemistry \cite{geckos, vdW-Chemistry} to physics, engineering, and
nanotechnology \cite{Milton04, Lamoreaux05, CasimirLivro, MostepanenkoRMP, MostepanenkoLivro}.

Recently, great  attention has been devoted to dispersive interactions in carbon
nanostructures, such as graphene sheets. These systems are specially appealing
since graphene possesses unique mechanical, electrical, and optical properties
\cite{Graphene}. Recently, dispersive interactions between graphene sheets
and/or material planes have been investigated~\cite{Bordag2009,Bordag2012,Drosdoff,Fialkovsky,Galina2013,Gomez,Macdonald_PRL,Sernelius,Svetovoy, Klimchitskaya-Mostepanenko-14}, as well as the Casimir-Polder interaction between
atoms and graphene \cite{Chaichia,Churkin,Galina2014,Judd,Sofia-Scheel-13}. In particular, the impact of
graphene coating on the atom-plate interaction has been calculated for different
atomic species and substrates; in some cases this results in modifications of
the order of $20\%$ in the strength of the
interaction~\cite{Galina2014}. Furthermore, results on the
possibility of shielding the dispersive interaction with the aid of graphene
sheets have been reported~\cite{Sofia-Scheel-13}.
\begin{figure}
\vspace{0.1in}
  \centering
  \includegraphics[scale = 0.34]{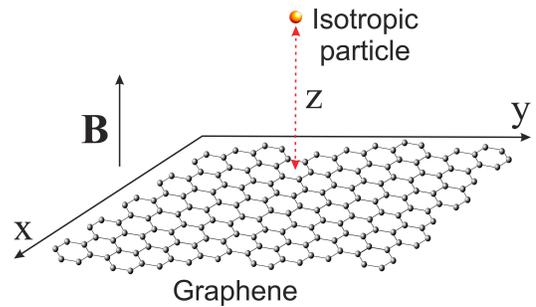}
  \caption{System of suspended graphene and an isotropic particle at a distance $z$, in presence of a perpendicular static magnetic field.}
  \label{Figure1}
\end{figure}

However, the possibility of controlling the Casimir-Polder interaction
between an atom and a graphene sheet by an external agent has never been
envisaged so far. The possibility of varying the atom-graphene interaction
without changing the physical system would be extremely appealing for both
experiments and applications. With this motivation, and
exploring the extraordinary magneto-optical response of
graphene, in the present work we investigate the Casimir-Polder interaction
between a Rubdium atom and a suspended graphene sheet under the influence of an
external magnetic field ${\bf B}$. We show that just by changing the applied
magnetic field, the atom-graphene interaction can be greatly reduced, up to
$80\%$  of its value without the field. Furthermore, we demonstrate that
at low temperatures ($T\simeq 4$ K) the Casimir-Polder energy exhibits sharp
discontinuities at certain values of $B$, that we show to be a manifestation of
the quantum Hall effect. As the distance $z$ between the atom and the graphene
sheet grows to $z \gtrsim 1 \ \mu$m, these discontinuities form a
plateau-like pattern with quantized values for the Casimir-Polder energy.  We
also show that at room temperature ($T\simeq 300$ K) thermal effects must be
taken into account even for considerably short distances. Moreover, in this
case the discontinuities in the atom-graphene interaction do not
occur, although the Casimir-Polder energy can still be tuned in $\sim 50 \%$ by
applying an external magnetic field.

Let us consider that an isotropic particle is placed at a distance $z$ above a
suspended graphene sheet biased by a static magnetic field  ${\bf B} = B
\hat{{\bf z}}$, as depicted in Fig.~\ref{Figure1}. The whole system is assumed
to be in thermal equilibrium at temperature $T$. When $B \neq 0$ the
optical properties of graphene can be well described in terms of a homogeneous
but anisotropic two-dimensional conductivity tensor. In this case, the
Casimir-Polder (CP) energy interaction can be calculated trough the scattering
approach and can be cast as \cite{Paulo1}
\begin{eqnarray}
U_T(z)&=&\frac{k_B T}{\varepsilon_0 c^2}{\sum_{l=0}^\infty}' \xi_l^2 \,\alpha(i\xi_l)\int\frac{d^2\textbf{k}}{(2\pi)^2}\frac{e^{-2\kappa_l z}}{2\kappa_l}  \cr\cr
&& \hspace{-40pt}\times \Bigg[ r^{s,s}(\textbf{k},i\xi_l, B)- \Bigg(1+\frac{2c^2k^2}{\xi_l^2}\Bigg)r^{p,p}(\textbf{k},i\xi_l, B)\Bigg]\, ,
\label{uT}
\end{eqnarray}
where $\xi_l = 2 \pi l / (\hbar k_B T)$ are the so called Matsubara frequencies,
$\kappa_n=\sqrt{\xi_l^2/c^2 + k^2}$, $\alpha(i\xi)$ is the electric
polarizability of the particle, and $r^{s,s}(\textbf{k},i\xi, B)$,
$r^{p,p}(\textbf{k},i\xi, B)$ are the diagonal reflection coefficients
associated to graphene. As usual, the prime in the summation means that the
zero-th term has to be weighted by a factor of 1/2. Note that the
cross-polarization reflection coefficients $r^{s,p}(\textbf{k},i\xi, B)$ and
$r^{p,s}(\textbf{k},i\xi, B)$, despite being non-vanishing, do not appear in
(\ref{uT}). This, however, does not mean that anisotropy plays no
role in the interaction energy, as transverse conductivities/permittivities
could still appear in $r^{s,s}(\textbf{k},i\xi, B)$, $r^{p,p}(\textbf{k},i\xi,
B)$. In particular, by modeling graphene as a $2$-D material with a surface
density current ${\bf K} = \mbox{{\mathversion{bold}${\sigma}$}} \cdot {\bf
E}|_{z=0}$ and applying the appropriate boundary conditions to the
electromagnetic field, one can show that the reflection coefficients
are~\cite{Macdonald_PRL}
\begin{eqnarray}
&& \hspace{0pt}r^{s,s}({\bf k},i\xi,B) = \nonumber \\
&& \hspace{10pt} \dfrac{2\sigma_{xx}(i\xi, B) Z^h +
\eta_0^2[\sigma_{xx}(i\xi, B)^2+\sigma_{xy}(i\xi, B)^2]}
{-\Delta({\bf k},i\xi, B)}\, ,  \\
&& \hspace{0pt} r^{p,p}({\bf k},i\xi,B)= \nonumber \\
&& \hspace{10pt}\dfrac{2\sigma_{xx}(i\xi, B) Z^e +
\eta_0^2[\sigma_{xx}(i\xi, B)^2+\sigma_{xy}(i\xi, B)^2]}
{\Delta({\bf k},i\xi, B)}\, ,  \\
&&\Delta({\bf k},i\xi, B) = [2 + Z^h \sigma_{xx}(i\xi, B)][2
+ Z^e \sigma_{xx}(i\xi,B)] \nonumber \\
&& \hspace{10pt} +[\eta_0\sigma_{xy}(i\xi, B)]^2  \, ,
\label{RefCoefs}
\end{eqnarray}
where $Z^h = \xi \mu_0 / \kappa$, $Z^e = \kappa/(\xi \epsilon_0)$, and
$\eta_0^2=\mu_0/\epsilon_0$. Besides, $\sigma_{xx}(i\xi, B)$ and
$\sigma_{xy}(i\xi, B)$ are the longitudinal and transverse conductivities of
graphene, respectively.

The electric conductivity tensor of graphene under an external magnetic field is
well known and reads \cite{Gusynin}
\begin{widetext}
\vskip -0.4cm
\begin{eqnarray}
\sigma_{xx}(i\xi,B) &=& \dfrac{e^3v_F^2 B\hbar(\xi+\tau^{-1})}{\pi c} \sum\limits_{n=0}^\infty \Bigg\{\dfrac{n_F(M_n)-n_F(M_{n+1})+n_F(-M_{n+1})-n_F(-M_n)}    {D_n(M_{n+1}-M_n)}
+(M_n\to-M_n)\Bigg\}\, , \label{Conductivity1}\\
\sigma_{xy}(i\xi,B)&=&\dfrac{e^3v_F^2B}{-\pi c}\sum\limits_{n=0}^\infty \{n_F(M_n)-n_F(M_{n+1})-n_F(-M_{n+1})+n_F(-M_n)\} \Bigg[\dfrac{1}{D_n}
+(M_n\to-M_n) \Bigg]\, ,
\label{Conductivity2}
\end{eqnarray}
\end{widetext}
where $1/\tau$ is a phenomenological scattering rate, $n_F(E)$ is the
Fermi-Dirac distribution, $D_n = (M_{n+1}-M_n)^2+\hbar^2(\xi+\tau^{-1})^2$,
$M_n=\sqrt {n} M_1$ are the Landau energy levels, $M_1^2 = 2 \hbar e B v_F^2/c$
is the Landau energy scale and $v_F \simeq 10^6$ m/s is the Fermi velocity.
Also, in the following we use $\tau = 1.84 \times 10^{-13}$ s and set the
chemical potential at $\mu_c = 0.115$ eV.

We still have to specify the particle in our setup. It turns out that a Rubidium
atom is a convenient choice, since there are experimental data on its complex
electric polarizability $\alpha(\omega)$ for a wide range of frequencies
\cite{Rubidium}. As it is clear from (\ref{uT}), we actually need the
polarizability evaluated at imaginary frequencies, which can be readily be
obtained from the Kramers-Kronig relations provided one has the data for ${\rm
Im} \, \alpha(\omega)$ \cite{Paulo1}.
%
%
\begin{figure}[h]
  \centering
  \includegraphics[scale = 0.45]{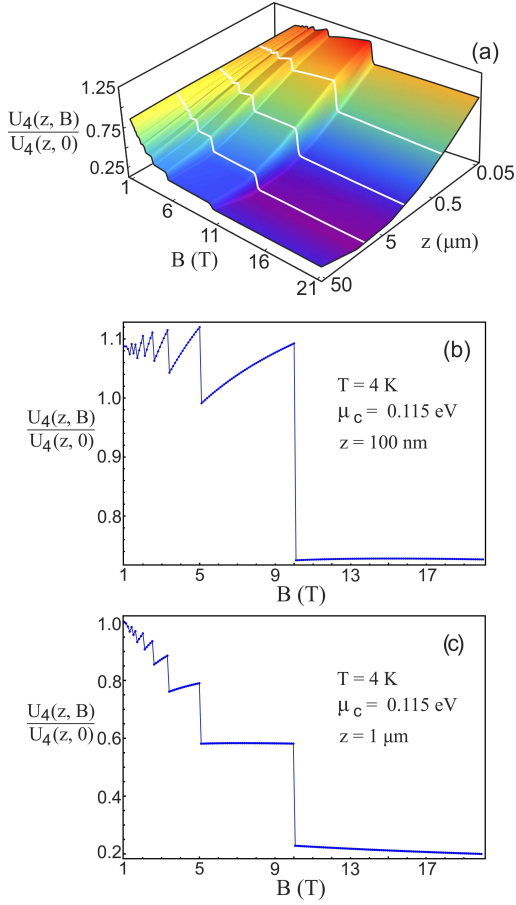}\\
  \caption{{\bf Above:} The Casimir-Polder energy of a Rubidium atom in front of
a graphene sheet subjected to a magnetic field $B$, as a function of $B$ and the distance $z$.
{\bf Below:} The Casimir-Polder energy as a function of $B$ for two fixed distances: (b) $z = 100$ nm, and (c) $z = 1$ $\mu$m.
In all plots $\mu_c = 0.115$ eV, $T = 4$ K and we have normalized $U_4(z,B)$ by the energy in the absence of magnetic field $U_4(z,0)$.}
\label{Figure2}
\end{figure}

\begin{figure}
\vspace{0.1in}
  \centering
  \includegraphics[scale = 0.35]{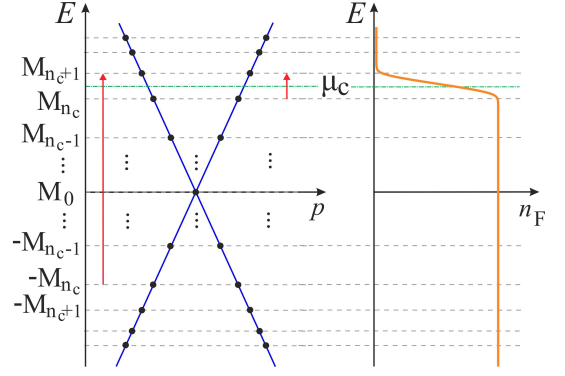}
  \caption{{\bf Left:} energy-momentum dispersion diagram of graphene in a magnetic field. The blue lines show the usual linear dispersion relation of graphene whilst the Landau levels brought forth by the introduction of $B$ are represented by the black dots. The long (short) vertical arrow shows the lowest energy interband (intraband) transition crossing the chemical potential (dot-dashed green line). {\bf Right:} Fermi-Dirac distribution at temperature $T$.}
    \label{Figure3}
\end{figure}

Our first results are summarized in Fig. \ref{Figure2}a, where we depict the CP
energy of our setup as a function of the atom-graphene distance $z$ and magnetic
field $B$ for $T = 4$ K (normalized by its corresponding
value for $B=0$ T). The hallmark of this plot is, surely, the great amount of
discontinuities shown by the energy as a function of $B$ for all distances
considered. These drops show up even more
clearly in Figs. \ref{Figure2}b and \ref{Figure2}c, where we take two cuts of
Fig.~\ref{Figure2}a at two different fixed values of $z$, namely $z=100$ nm and
$z=1$ $\mu$m, and present them as 2-D plots.  Such discontinuities are directly
linked to the discrete Landau levels brought about by the application
of a magnetic field. In order to understand the situation, let us consider the
energy-momentum dispersion diagram of graphene in the presence of a static
magnetic field depicted in Fig.~\ref{Figure3}. On the left the usual linear
dispersion relation of graphene is presented by the blue solid line. Due to the
magnetic field the carriers in graphene can occupy only the
discrete values of energy, given by the Landau levels $M_n$ represented
by black dots. The allowed transitions between two Landau levels give rise
to all terms of the summations in Eqs. (\ref{Conductivity1}) and
(\ref{Conductivity2}). There are two kinds of transitions: interband
transitions, that connect levels at distinct bands ({\it e. g.} long arrow
between $-M_{n_c}$ and $M_{n_c+1}$), and intraband transitions that involve
levels at the same band ({\it e. g.} short arrow between $M_{n_c}$ and
$M_{n_c+1}$).
The possibility of occurrence of a specific transition is related to the
difference between the probabilities of having the initial and final levels
full and empty, respectively. Ultimately, these probabilities are given by the
Fermi-Dirac distribution (solid orange line on the right of Fig. \ref{Figure3}).
Hence whenever a given Landau level, whose position in energy depends
on $B$, crosses upwards (downwards) the chemical potential (dot-dashed green
line) of the graphene sheet, it gets immediately depopulated (populated) as a
consequence of the quasi-step-function character of the Fermi-Dirac distribution
at 4 K. Therefore the crossing of the $n$-th level
sharply quenches the $M_{n} \leftrightarrow M_{n+1}$ transition; at the same time that it gives birth to the $M_{n-1}
\leftrightarrow M_{n}$ one \cite{Gusynin}, in a process that changes the
conductivity, and thus the interaction energy, discontinuously. The fact that
the CP energy always drops {\it down} at a discontinuity as we increase $B$ may
be understood by recalling the behavior of the relativistic Landau levels with
$\sqrt{n}$ (see above). This square-root growth implies that the $n-1$
$\leftrightarrow$ $n$ gap is wider than the $n$ $\leftrightarrow$ $n+1$ one,
making the transition weaker, hence reducing the overall conductivity.
A similar analysis is valid for the interband transitions.

Figure \ref{Figure2} also reveals that a flattening of the steps in the
CP energy between drops occurs as $B$ increases. However, if on the one hand for
$z=100$ nm only the very last step is really flat, on the other hand many
plateau-like steps exist for $z=1$ $\mu$m. This result is connected to the
electrostatic limit of the conductivity: for large distances the
exponential factor in (\ref{uT}) strongly suppress all contributions coming from
$l \neq 0$, whereas for $l=0$ and large magnetic fields $\sigma_{xx}
\sim 0$ and $\sigma_{xy} \sim \pm (2N+1)e^2/(\pi\hbar)$, where $N$ is an integer
\cite{Graphene}. Therefore, in the limit of large
distances (of the order of micrometers) only the Hall conductivity contributes
to $U_{4}(z, B)$, and the CP energy became almost quantized~\cite{Macdonald_PRL}. Furthermore, one should note the
striking reduction in the force as we sweep through different values of $B$.
While for $z = 100$ nm this reduction can be as hight as $30\%$, one can get up
to an impressive $80\%$ decrease in the CP interaction for $z=1$ $\mu$m
and  $B \gtrsim$ 10 T, with huge drops in between. Finally, it should be
remarked that for $B \gtrsim$ 10 T the CP interaction is practically insensitive
to changes in the magnetic field, regardless of the atom-graphene
distance. In this regime the discontinuities in the CP energy do not
occur any longer. This effect has its origins in the fact that there is
a critical value of the magnetic field $B_c$ (in the present case, $B_c \sim 10$
T) for which the transition $M_0 \rightarrow M_1$ is dominant since all
Landau levels, except $M_0$, are above the chemical potential.
Altogether, our findings suggest that the atom-graphene is a
particularly suited system for investigation of the effects of external magnetic
fields on CP forces, and may pave the way for an active modulation of dispersion
forces in general.
\begin{figure}[!h]
\vspace{0.1in}
  \centering
  \includegraphics[scale = 0.35]{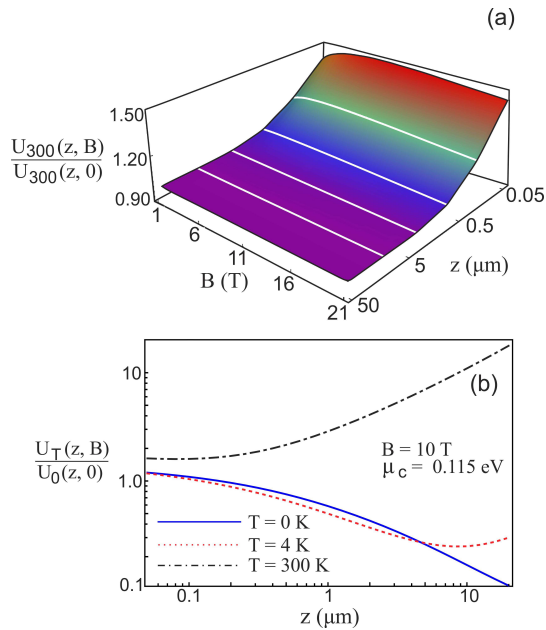}
  \caption{{\bf Above:} the Casimir-Polder interaction energy $U_{300}(z,B)$
(normalized by $U_{300}(z,0)$) between a Rubidium atom and a graphene sheet as a
function of both distance and external magnetic field strength for $T = 300$ K.
{\bf Below:} the Casimir-Polder energy (normalized by $U_{0}(z,0)$) as a
function of the mutual distance between atom and graphene sheet for $T = 0$ K
(solid line), $T = 4$ K (dashed line), and $T = 300 K$ (dot-dashed line). In
both panels (a) and (b) the graphene chemical potential is $\mu_c = 0.115$ eV.}
    \label{Figure4}
\end{figure}

In order to investigate thermal effects, Fig.~\ref{Figure3}a we present
the CP energy as a function of both $z$ and $B$ at room temperature. The most
distinctive aspect of Fig.~\ref{Figure3}a is the complete absence of
discontinuities that characterize the behavior of the CP energy at low
temperatures. At $T = 300$ K the Fermi-Dirac distribution is a quite smooth
function of the energy levels, allowing for a partial filling of many Landau
levels. Hence the effects of the crossing between these levels and the
graphene's chemical potential is hardly noticed, resulting in a smooth CP energy
profile. Another important aspect of Fig.~\ref{Figure3}a is that the CP energy
becomes essentially independent of $B$ for $z \gtrsim 1$ $\mu$m. For this set of
parameters, the system is already in the thermal regime,
where the CP energy is essentially dominated by the electrostatic conductivity.
In this regime, the CP energy is very weakly affected by variations in $B$ due
the already discussed exponential suppression of the $l \geq 1$  terms in Eq.
(\ref{uT}). We emphasize, however, that the absence of discontinuities does not
prevent one from tuning the CP interaction between a Rb atom and a graphene
sheet at least at short distances. This tunability can be achieved even for
relatively modest magnetic fields, as the value of the CP energy can increase up
to 50 $\%$ (compared to the case where $B = 0$ T) by applying a magnetic field
of $B=5$ T for $z = 50$ nm. For $B=5$ T and $z = 100$ nm, the variation in the
interaction can still be as high as 30$\%$. In Fig.~\ref{Figure3}b the CP energy
is calculated for $B=10$ T and for different temperature values, $T = 0$, $4$,
and $300$ K, all normalized by the zero-temperature, zero-field energy value
$U_0(z,0)$. Figure~\ref{Figure3}b reveals that thermal corrections are relevant
even for low temperatures, and for a broad range of distances: we have
a 10-20\% variation in the relative difference of $U_4(z,10)$ and
$U_0(z,10)$ in the 1-10 $\mu$m interval, which is in the ballpark of
recent/current experiments' precision. Besides, Fig.~\ref{Figure3}b
demonstrates that at room temperature not only the thermal effects are
absolutely dominant in the micrometer range, but they also play an important
role even for small distances. Indeed, at $z = 100$ nm the relative difference
between $U_{300}(z,10)$ and $U_0(z,10)$ is $\sim 45\%$ and at $z = 1$ $\mu$m
it is $\sim 400 \%$; so in the latter approximately $80 \%$ of the CP
energy come from thermal contribution. We conclude that, at room
temperature, these effects should be taken into account for a wide range of
distances between the atom and the graphene sheet.

In conclusion, we have investigated the dispersive Casimir-Polder interaction
between a Rubidium atom and a suspended graphene sheet subjected to an external
magnetic field ${\bf B}$. Apart from providing a concrete physical system where
the dispersive interaction in nano and micrometer scales can be controlled by an
external agent, we show that just by changing the applied magnetic field, this
interaction can be reduced up to $80\%$  of its value in the absence of the
field. Further, due to the quantum Hall effect, we show that for low
temperatures the Casimir-Polder interaction energy acquires sharp
discontinuities at given values of $B$ and that these discontinuities approach
a plateau-like pattern with a quantized Casimir-Polder interaction energy as the
atom and the graphene sheet become more and more far apart. In addition, we
show that at room temperature thermal effects must be taken into account even
for considerably short distances. In this case, the discontinuities in the
atom-graphene dispersive interaction are not present any longer, although the
interaction can still be tuned in $\sim 50 \%$ by applying an external magnetic
field.

We thank P. A. Maia Neto for assistance with the Rubidium data, and R. S.
Decca, I. V. Fialkovsky, E. C. Marino, and N. M. R. Peres for useful discussions. We also
acknowledge CNPq and FAPERJ for partial financial support.


\end{document}